\documentclass[aip,jcp,amsmath,amssymb,preprint,longbibliography]{revtex4-2}

\usepackage{graphicx}
\usepackage{hyperref}
\usepackage{color}
\usepackage{ulem}

\def\gs0{\gamma_\mathrm{S0}}
\def\ptl{\partial}
\def\ds{\displaystyle}
\def\ie{\textit{i.e.}, }
\def\eg{\textit{e.g.}, }
\def\angb#1{\left<#1\right>}
\def\bm#1{\mbox{\boldmath $#1$}}
\def\dint{\int\!\!\!\int}
\def\tint{\int\!\!\!\int\!\!\!\int}
\def\rd{\mathrm{d}}

\def\eqref#1{(\ref{#1})}
\newcommand{\osaka}{Department of Mechanical Engineering, Osaka University, 2-1 Yamadaoka, Suita 565-0871, Japan}
\newcommand{\osakacu}{Department of Mechanical and Physical Engineering, Osaka City University, 
3-3-138 Sugimoto,  Sumiyoshi-ku, Osaka 558-8585, Japan}
\newcommand{\tuswater}{Water Frontier Research Center (WaTUS),
Research Institute for Science \& Technology,
Tokyo University of Science,
1-3 Kagurazaka, Shinjuku-ku, Tokyo, 162-8601, Japan}
\begin{document}

\begin{flushright}
\scalebox{0.8}{
The following article has been submitted to \textit{The Journal of Chemical Physics}.} 
\scalebox{0.8}{\copyright 2021 Hiroki Kusudo. This article is distributed under a Creative Commons Attribution (CC BY) License.}
\end{flushright}


\title{
Local stress tensor calculation by the Method-of-Plane in microscopic systems with macroscopic flow: a formulation based on the velocity distribution function
}

%
\author{Hiroki Kusudo}
\email{hiroki@nnfm.mech.eng.osaka-u.ac.jp}
\affiliation{\osaka}
\author{Takeshi Omori}
\email{omori@osaka-cu.ac.jp}
\affiliation{\osakacu}
\author{Yasutaka Yamaguchi}
\email{yamaguchi@mech.eng.osaka-u.ac.jp}
\affiliation{\osaka}
\affiliation{\tuswater}
\date{\today}

\begin{abstract}
In this work,  we showed a calculation method of local stress tensor applicable to non-equilibrium MD systems based on the Method of Plane (MoP).
From the relation between the macroscopic velocity distribution function and the microscopic molecular passage across a fixed control plane, 
we derived a method to calculate the basic properties of the macroscopic momentum conservation law including the density, the velocity, the momentum flux, the interaction and kinetic terms of the stress tensor defined on a surface with a finite area. 
Any component of the streaming velocity can be obtained on a control surface, which enables the separation of the kinetic momentum flux into the advection and stress terms in the framework of MoP.
We verified the present method through the extraction of the density, velocity and stress distributions in a quasi-1D steady-state Couette flow system and in a quasi-2D steady-state system with a moving contact line. 
In our method, as opposed to volume average method, the density, mass and momentum fluxes are defined on a surface, which is essential to be consistent with the mass and momentum conservation laws 
in dynamic systems.

\end{abstract}


\maketitle 

%
\section{Introduction}
\label{sec:Intro}
With the increasing interest in microfluidic devices and nanotechnologies, 
molecular dynamics (MD) simulations have become a powerful computational 
tool to examine the fluid behavior for small scale systems.
For the understanding in terms of flow fields, the microscopic motion of individual molecules must be averaged, and the stress tensor plays a key role in such macroscopic flow. 
Within the framework of fluid mechanics, the stress tensor is determined from the  velocity fields through the constitutive equation typically including the viscosity, and 
the local acceleration of the fluid is given by the gradient of the 
stress tensor as well as the external field to satisfy the momentum 
conservation. On the other hand, the molecular motion is governed by 
the intermolecular interaction, and the stress tensor should be defined 
through the average of the molecular motion and interaction.
\par
%
%
\def\Jdkin#1{\dot{J}^\mathrm{kin}_{#1}}
\par
More concretely, the macroscopic equation of continuity is given 
with the density $\rho(\bm{x},t)$ and velocity vector $\bm{u}(\bm{x},t)$ 
both as functions of position $\bm{x}$ and time $t$ by
\begin{align}
    \frac{\ptl \rho }{\ptl t} 
    + 
    \frac{\ptl \rho u_{k}}{\ptl x_{k}}
    =
    0,
    \label{eq:mass_conserv_macro}
\end{align}
where $x_{k}$ and $u_{k}$ are the $k$-direction components of 
the position and the velocity vector, respectively.
The Einstein notation is used with a dummy 
index $k$ for the vectors. The second term of the LHS is called the advection or streaming term.
Equation~\eqref{eq:mass_conserv_macro} comes from the 
mass conservation
\begin{align}
    \tint_{V}  \rd V \frac{\ptl \rho}{\ptl t}
    = -\dint_{S} \rd S \rho u_{k}  n_{k}
    \label{eq:mass_conserv_macro_vol_surf}
\end{align}
satisfied for an arbitrary volume $V$ in an enclosing surface $S$, 
where $n_{k}$ is the $k$-direction component of the outward 
unit normal vector $\bm{n}$ with respect to the infinitesimal surface element $\rd S$. 
This integral form in Eq.~\eqref{eq:mass_conserv_macro_vol_surf}
means that the mass in $V$ can be changed only by the mass 
flux passing the surface $S$. 
By applying Gauss' divergence theorem for the RHS as
\begin{align}
    \dint_{S} \rd S \rho u_{k}  n_{k}
    = \tint_{V} \rd V \frac{\ptl \rho u_{k}}{\ptl x_{k}}, 
    \label{eq:mass_gauss}
\end{align}
it follows for Eq.~\eqref{eq:mass_conserv_macro_vol_surf} that 
\begin{align}
    \tint_{V}  \rd V \left(
    \frac{\ptl \rho}{\ptl t}
    +
    \frac{\ptl \rho u_{k}}{\ptl x_{k}}\right) = 0,
    \label{eq:mass_conserv_macro_volint}
\end{align}
for an arbitrary $V$, which corresponds to Eq.~\eqref{eq:mass_conserv_macro}. 
\par
Similarly, the macroscopic momentum equation, 
the Navier-Stokes equation, is given by
\begin{align}
    \frac{\ptl \rho u_{l}}{\ptl t} 
    + 
    \frac{\ptl \rho u_{l}u_{k}}{\ptl x_{k}}
    =
    \frac{\ptl \tau_{kl}}{\ptl x_{k}}
    +
    \rho F_{l}, 
    \label{eq:NS_macro}
\end{align}
where the fluid stress tensor component $\tau_{kl}$ expresses 
the stress in the $l$-direction exerted on a surface element 
with an outward normal in the $k$-direction, and $F_{l}$ denotes
the external force per mass. This is also derived from
the momentum conservation for an arbitrary volume $V$ enclosed by $S$:
%
\begin{align}
    \tint_{V}  \rd V \frac{\ptl \rho u_{l}}{\ptl t}
    = 
    -
    \dint_{S} \rd S \rho u_{l} u_{k} n_{k}
    +
    \dint_{S} \rd S \tau_{kl} n_{k}
    +
    \tint_{V}  \rd V \rho F_{l},
    \label{eq:mom_conserv_macro_vol_surf}
\end{align}
%
meaning that the total momentum in $V$ can be changed by the 
momentum flux passing the surface $S$ as well as the 
impulse due to the stress exerted on the surface $S$ 
and external force exerted on the volume $V$. 
Specifically 
note that the advection $\rho u_{l}u_{k}$ and stress $\tau_{kl}$ in the 1st and 2nd terms of the RHS,
respectively, are separated.
\par
In contrast to the above-mentioned macroscopic feature, 
microscopic molecules have their own velocities, and their 
local average and variance correspond to the macroscopic 
velocity $\bm{u}$ and temperature under the assumption 
of local equilibrium in space and time, respectively. 
Hence, an instantaneous microscopic mass or momentum transfer across an arbitrary surface $S$ exists due to the passage of the constituent molecules even in macroscopically static system without mean flow with $\bm{u}=\bm{0}$. 
In the kinetic theory of gases based on the Boltzmann equation with respect to the velocity distribution function (VDF) of the constituent molecules, 
the macroscopic velocity is subtracted from the microscopic molecular velocity upon the definition of the stress. 
Before the establishment of the MD method, \Citet{Irving1950} (IK)
put forward the connection between the macroscopic conservation 
laws and microscopic molecular motion governed by the 
inter-molecular interaction through the statistical 
mechanical theory based on the distribution function in the phase space, 
and derived an expression of the local pointwise
stress comprised of kinetic and interaction parts
including Taylor series expansion of differences in delta 
functions to express the microscopic particle feature.
\par
After the introduction of numerical MD simulations,~\cite{Alder1957,Alder1959} the calculation 
of average stress in homogeneous bulk systems, equivalent to the average bulk pressure with its sign 
inverted, was enabled based on the virial theorem, which
indeed corresponds to the average of the IK form 
integrated in space and time.
From this bulk stress, the viscosity as a transport coefficient can 
also be obtained based on the Green-Kubo relation from 
the time-fluctuation of the off-diagonal stress component 
averaged in space.~\cite{Evans2008,Allen2017}
Regarding the local stress implemented for MD simulations 
with a discrete time-step, \citet{Tsai1979} 
proposed a pragmatic scheme to calculate the averaged stress
defined on a flat plane in quasi-1D planer systems, and 
\citet{Thompson1984} extended the approach toward a spherical 
curved surface for the analysis of the surface tension of 
a spherical droplet. In both cases, all the momentum flux 
and intermolecular force across the plane or the sphere, 
which divide the computational domain, were summed up 
during the time integration in macroscopically static systems. This type of stress definition is usually called the Method 
of Plane (MoP),~\cite{Todd1995} or 
\citet{Hardy1982} stress, where  
the momentum conservation law
with the above-mentioned MoP for quasi-1D systems was proved to be an exact consequences of Newton's laws,~\cite{Hardy1982} \ie the MoP meets Gauss' divergence theorem for 
any control volume (CV) surrounded by enclosing surface(s) irrespective of whether the local system in the control volume is homogeneous or not. In addition, this momentum conservation is not restricted to quasi-1D systems but also applicable 
to quasi-2D systems, \eg the present authors adopted the MoP for a CV with an rectangular enclosing surface set around the contact line of an equilibrium droplet to examine the nanoscale wetting behavior through the mechanical balance exerted on the fluid in the CV.~\cite{Yamaguchi2019}
%
On the other hand, the volume average (VA) is proposed 
as another approach that gives the local mean stress in space, 
where weighted average of the pair interaction in local CVs is included in the formulation.~\cite{Lutsko1988,Weng2000,Cormier2001,Heyes2011,Yang2012,Heyes2016,Smith2017,Shi2021,Smith2020}
This VA should in principle be applied only for homogeneous CVs, and is advantageous especially to explore the link with the local  fluctuation and thermodynamic limit,~\cite{Heyes2016,Smith2017} while stress integral can also be obtained by the VA, and this enables the calculation of the surface tension based on Bakker's equation using this stress integral,~\cite{Bakker1928}
which is known as a mechanical route to evaluate the surface tension.~\cite{Rowlinson1982,Weng2000,Nishida2014,Surblys2014,Allen2017,Smith2020} 
The momentum conservation is satisfied for the whole system if the VA is properly summed up; however,  special care is needed to consider the momentum conservation for local CVs because the VA originally was not designed to examine local momentum conservation to be satisfied through the link to Gauss' divergence theorem.~\cite{Hardy1982,Heyes2011,Shi2021} This feature is similar to the atomic stress,~\cite{Thompson2009} for instance provided as \textit{stress/atom} command in LAMMPS package,~\cite{LAMMPS1995} often used to simply visualize the stress field.
\par
Going back to the momentum conservation in Eq.~\eqref{eq:mom_conserv_macro_vol_surf}, 
for systems with a non-zero local flow, \ie macroscopically dynamic systems with $\bm{u}\neq \bm{0}$, the local macroscopic velocity $\bm{u}$ must be defined on 
a surface $S$ enclosing a control volume $V$ 
in MD systems so that the stress calculated in MD simulations may be consistent with the macroscopic momentum Eq.~\eqref{eq:NS_macro}. In other words, the momentum transfer due to the microscopic molecular passage across a control surface should be separated into advection and stress contributions to examine the local flow from a macroscopic point of view.
%
%
%
%
\par
In this paper, we show a calculation method of the MoP-based local stress tensor applicable to non-equilibrium molecular dynamics (NEMD) systems. We provide the formulation for systems consisting of single-component mono-atomic fluid molecules for simplicity while the present framework 
is also applicable to systems of multi-component or poly-atomic fluid molecules. 
For the derivation, we introduced the VDF to clearly give the average of physical properties defined on a fixed control plane 
as we shall see later. With this procedure, we provide not only the expression of the kinetic term of stress tensor but also those of the components in the advection term, \ie the density and macroscopic velocity as an extension of the MoP.
To check its validity, we performed test calculation in two systems: 1) a quasi-1D Couette flow system and 2) a quasi-2D system with  liquid-solid-vapor contact lines, both consisted of a Lennard-Jones fluid between parallel solid walls moving in the opposite directions tangential to the walls. In the first system, we compared the density and velocity distributions obtained by the present method and the VA, and we calculated the distributions of the stress components and advection term. Furthermore,
we showed that the same velocity distribution was obtained on bin faces with different normal directions, which is essential to determine the advection term. 
In the second system, the density, velocity and stress distributions are calculated in the complex flow with 
liquid-vapor interfaces and contact lines.
%
\section{Theory}
%
We show the derivation of the stress averaged on a finite bin face in a Cartesian coordinate system for single-component mono-atomic fluid in the following for simplicity.
Note that the Einstein notation with dummy indices used in Sec.~\ref{sec:Intro} is not applied hereafter.
The fluid stress tensor component $\tau_{kl}$ 
which expresses the stress in the $l$-direction exerted on a surface element with an outward normal in the $k$-direction, is given by the kinetic term $\tau_{kl}^\mathrm{kin}$ and the inter-molecular interaction term $\tau_{kl}^\mathrm{int}$ as %
\begin{align}
     \tau_{kl} = 
     \tau_{kl}^\mathrm{kin} +
     \tau_{kl}^\mathrm{int}.
     \label{eq:tau=taukin+tauint}
\end{align}
In the standard MoP for equilibrium MD systems 
without mean flow consisting of single-component 
mono-atomic fluid molecules,~\cite{Todd1995,Yamaguchi2019} 
the kinetic term 
$\tau_{kl}^\mathrm{kin}$ 
in Eq.~\eqref{eq:tau=taukin+tauint} 
on a bin face of area $S_{k}$ with its normal vector
pointing to the $k$-th Cartesian direction
is calculated by
\begin{equation}
\tau_{kl}^\mathrm{kin}
\equiv 
    -\frac{1}{S_{k}\delta t}
    \angb{
    \sum_{i\in\mathrm{fluid}, 
    \delta t}^{\text{across\ } S_{k}}
    m^{i} v_{l}^{i} \frac{v_{k}^{i}}{\left| v_{k}^{i}\right|} 
    },
    \label{eq:MoP_kin}
\end{equation}
where $m^{i}$ and $v_{l}^{i}$ denote the mass and $l$-component of 
the velocity vector $\bm{v}^{i}$ 
of fluid particle $i$, respectively. We also 
denote the bin face by $S_{k}$ hereafter. 
The angular brackets denote the ensemble average, and 
the summation
$\sum_{i\in\mathrm{fluid}, \delta t}^{\text{across\ } S_{k}}
$
is taken for every fluid particle $i$ 
passing through $S_{k}$ within a 
time interval of $\delta t$, which is equal to
the time increment for the numerical integration. 
Considering that we deal with a single-component fluid 
molecules of an identical mass $m$, we substitute 
$m^{i}$ with $m$ hereafter.
A sign function $\frac{v_{k}^{i}}{|v_{k}^{i}|}$ 
equal to $\pm 1$ is multiplied to the momentum transfer 
$mv_{l}^{i}$ across $S_{k}$ to evaluate the 
kinetic effect on the stress depending on the passing
direction. 
%
Note that in static equilibrium systems, \ie systems without macroscopic local mean flow, the advection term is zero 
%
%
in the whole system.
\par
On the other hand, 
the intermolecular interaction term
$\tau_{kl}^\mathrm{int}$ in Eq.~\eqref{eq:tau=taukin+tauint} 
in the case of simple two body potential is calculated by
\begin{align}
    \tau_{kl}^\mathrm{int}=
    -\frac{1}{S_{k}}
    \angb{
    \sum_{(i,j)\in \mathrm{fluid}}^{\text{across\ } S_{k}} 
    F^{ij}_{l} \frac{r^{ij}_{k}}{|r^{ij}_{k}|}
    },
    \label{eq:stress_int}
\end{align}
where $r^{ij}_{k}$ and $F^{ij}_{l}$ 
denote the $k$-component 
of the relative position vector 
$\bm{r}^{ij} \equiv \bm{x}^{j} - \bm{x}^{i}$ and 
the $l$-component of the force vector 
$\bm{F}^{ij}$ on particle $j$ at position $\bm{x}^{j}$ from particle $i$ at 
position $\bm{x}^{i}$, respectively. 
The summation 
$\sum_{(i,j)\in \mathrm{fluid}}^{\text{across\ }S_{k}}$
is taken for all line segments of the inter-particle interaction between $\bm{x}^{i}$
and $\bm{x}^{j}$ which cross $S_{k}$.
A sign function $\frac{r^{ij}_{k}}{|r^{ij}_{k}|}$ is 
multiplied for this interaction term to evaluate the 
force effect depending on the force direction.
Note that technically the fluid-solid interaction
can also be included as $i$-$j$ pair in the summation $\sum_{(i,j)}^{\mathrm{across\ }S_{k}}$
in Eq.~\eqref{eq:stress_int}, but only the fluid-fluid interaction was taken into
account as the fluid stress, and fluid-solid contribution was considered as an external force field.~\cite{Yamaguchi2019,Kusudo2019,Imaizumi2020}
Also note that for multi-component systems or systems with poly-atomic molecules, difficulties mainly arise to treat the interaction force between different kind of molecules or the constraint force~\cite{Andersen1983} of the polyatomic molecules, where the interaction forces should be properly implemented into the stress calculation to satisfy the conservation laws.~\cite{Surblys2019}
%
%

To extend the standard MoP to steady-state NEMD systems with a non-zero
macroscopic mean local flow, 
the mean velocity should be properly subtracted 
from the kinetic term $\tau_{kl}^\mathrm{kin}$
in Eq.~\eqref{eq:MoP_kin}
so that the macroscopic momentum flux as the advection term due to the mean velocity $\bm{u}$ may be included not in the stress term but in the advection term 
within the macroscopic description of 
the momentum conservation, 
\ie in the Navier-Stokes equation~\eqref{eq:NS_macro}.
%
In the following, 
we provide a general framework 
to connect a microscopic variable $\xi^{i}$ of particles 
and a macroscopic field value $\xi(\bm{x},t)$ averaged on $S_{k}$
under non-zero mean velocity 
based on
the local VDF in the Cartesian $xyz$-coordinate system. 
\par
At first, we define the VDF $f(\bm{x}, \bm{v}, t)$
for the mass with a velocity $ \bm{v}=(v_x, v_y, v_z)$ at position $\bm{x} =(x, y, z)$ at time $t$, 
which gives the local density 
$\rho(\bm{x},t)$ by
\begin{align}
\rho(\bm{x},t)
= 
\int_{-\infty}^{\infty}\rd v_{x}
\int_{-\infty}^{\infty}\rd v_{y} \int_{-\infty}^{\infty}\rd v_{z} 
f(\bm{x},  \bm{v},t)
\equiv
\tint_{-\infty}^{\infty} \rd \bm{v}f(\bm{x},  \bm{v},t),
\label{eq:dens}
\end{align}
where we rewrite 
$\int_{-\infty}^{\infty}\rd v_{x}
\int_{-\infty}^{\infty}\rd v_{y} \int_{-\infty}^{\infty}\rd v_{z} 
$
by 
$
\tint_{-\infty}^{\infty} \rd \bm{v}
$.
%
Then a microscopic variable $\xi^{i}$ per mass of particle $i$ can be related to a corresponding macroscopic field variable $\xi(\bm{x},t)$ as
\begin{equation}
\lim_{\delta t \to 0}
\left<\sum_{i\in\mathrm{fluid},\delta t}
^{\text{crossing } S_{k}}m\xi^{i}\right> 
\equiv
\lim_{\delta t \to 0}
\tint_{-\infty}^{\infty}\rd  \bm{v}
\int_{0}^{\left| v_{k}\right|\delta t}\rd x_{k}  S_{k} 
f(\bm{x},  \bm{v},t)
\xi(\bm{x},t).
%
\label{eq:integral}
\end{equation}
The RHS denotes the integral weighted with VDF considering an oblique pillar of a 
base area $S_{k}$ and a height 
$|v_k|\delta t$
with its central axis parallel to $ \bm{v}$ 
, which is typically assumed upon the derivation of 
the equilibrium pressure in the kinetic theory of gases. 

With the limit $\delta t \to 0$, and by rewriting 
the average of $f(\bm{x}, \bm{v},t)$ and $\xi(\bm{x},t)$ 
in the oblique pillar
by $f(S_{k}, \bm{v},t)$ and $\xi(S_{k},t)$, 
respectively, the integral with respect to $x_{k}$
in the RHS of Eq.~\eqref{eq:integral} writes
\begin{equation}
\lim_{\delta t \to 0}
\int_{0}^{\left| v_{k}\right|\delta t}\rd x_{k}  S_{k} 
f(\bm{x},  \bm{v},t)
\xi(\bm{x},t)
=
\lim_{\delta t \to 0}
 S_{k} f(S_{k}, \bm{v})\xi(S_{k}) \left| v_{k}\right|\delta t,
\end{equation}
and it follows for Eq.~\eqref{eq:integral} that 
\begin{equation}
\lim_{\delta t \to 0}
\left<\sum_{i\in\mathrm{fluid},\delta t}^{\text{crossing }S_{k}}m\xi^{i} \right> 
=
\lim_{\delta t \to 0}
S_{k}\tint_{-\infty}^{\infty}\rd  \bm{v}
f(S_{k}, \bm{v},t)\xi(S_{k},t) \left| v_{k}\right|\delta t.
\end{equation}
Hence, by dividing both sides by $S_{k}\delta t$, 
\begin{equation}
\tint_{-\infty}^{\infty}\rd  \bm{v}
f(S_{k}, \bm{v},t)\xi(S_{k},t) \left| v_{k}\right|
=
\lim_{\delta t \to 0}
\frac{1}{S_{k}\delta t}
\left<\sum_{i\in\mathrm{fluid},\delta t}^{\text{crossing }S_{k}}m\xi^{i} \right> 
\label{eq:def_average_xi_on_Sk}
\end{equation}
is derived as a basic equation for the connection 
between the macroscopic field variable $\xi(S_{k},t)$ and microscopic
variable $\xi^{i}$ which belongs to the constituent particle $i$  
upon crossing $S_{k}$. 
\par
Now, we proceed to the expressions of the macroscopic 
field variables averaged on $S_{k}$. 
By substituting $\xi(S_{k},t)$ and $\xi^{i}$ in Eq.~\eqref{eq:def_average_xi_on_Sk}
with $\frac{1}{|v_{k}|}$ and $\frac{1}{|v^{i}_{k}|}$, respectively, and using Eq.~\eqref{eq:dens}, it follows
\begin{equation}
\rho(S_{k},t)=
\lim_{\delta t \to 0}
\frac{1}{S_{k}\delta t}\left<
\sum_{i\in\mathrm{fluid},\delta t}^{\text{crossing } S_{k}}
\frac{m}{\left| v_{k}^{i}\right|}
\right>,
\label{eq:densitysum}
\end{equation}
where $v_{k}^{i}$ denotes the velocity component 
in the $k$-direction of particle $i$. 
Similarly, regarding the macroscopic mass flux $\rho u_{l}$ 
given by
%
\begin{align}
\rho u_{l}(\bm{x},t)=&
\tint_{-\infty}^{\infty}\rd  \bm{v}
f(\bm{x}, \bm{v},t)v_{l},
\label{eq:massflux}
\end{align}
substituting $\xi(S_{k},t)$ and $\xi^{i}$ in Eq.~\eqref{eq:def_average_xi_on_Sk}
with
$\frac{v_{l}}{|v_{k}|}$ and $\frac{v_{l}^{i}}{|v_{k}^{i}|}$, respectively,
leads to
\begin{equation}
\rho u_{l} (S_{k},t)=
\lim_{\delta t \to 0}
\frac{1}{S_{k}\delta t}\left<
\sum_{i\in\mathrm{fluid},\delta t}^{\text{crossing } S_{k}}
\frac{mv_{l}^{i}}{\left| v_{k}^{i}\right|} 
\right>.
\label{eq:massfluxsum}
\end{equation}
From Eqs.~\eqref{eq:densitysum} and \eqref{eq:massfluxsum}, 
the macroscopic velocity $u_{l}$ results in 
\begin{equation}
u_{l}(S_{k},t)=\frac{\rho u_{l}(S_{k},t)}{\rho(S_{k},t)}
=
\lim_{\delta t \to 0}
\frac{
\ds\left<
\sum_{i\in\mathrm{fluid},\delta t}^{\text{crossing } S_{k}}
\frac{mv_{l}^{i}}{\left| v_{k}^{i}\right|} \right>
}{
\ds\left<
\sum_{i\in\mathrm{fluid},\delta t}^{\text{crossing } S_{k}}
\frac{m}{\left| v_{k}^{i}\right|}\right>
}
.\label{eq:averagedvelocity}
\end{equation}
\par
Finally, to write the kinetic contribution of the 
stress $\tau_{kl}^{\text{kin}}$, we use the expression 
in the kinetic theory of gases given by
\begin{align}
\tau_{kl}^{\text{kin}}(\bm{x},t)=-
\tint_{-\infty}^{\infty}\rd \bm{v}
f(\bm{x}, \bm{v},t)\left(v_{k}-u_{k}(\bm{x},t)\right)
\left(v_{l}-u_{l}(\bm{x},t)\right).
\label{eq:sigma_0}
\end{align}
By expanding Eq.~\eqref{eq:sigma_0}, it follows
\begin{align}
\nonumber
\tau_{kl}^{\text{kin}}(\bm{x},t)=&
-\tint_{-\infty}^{\infty}\rd \bm{v}
f(\bm{x}, \bm{v},t) v_{k}(v_{l}-u_{l}(\bm{x},t))  +u_{k}(\bm{x},t)\tint_{-\infty}^{\infty}\rd \bm{v} f(\bm{x}, \bm{v},t)(v_{l}-u_{l}(\bm{x},t))
\\ \nonumber
=&-\tint_{-\infty}^{\infty}\rd \bm{v}
f(\bm{x}, \bm{v},t)v_{k}(v_{l}-u_{l}(\bm{x},t)) 
+ u_{k}\rho u_{l}(\bm{x},t)-u_{k}\rho u_{l}(\bm{x},t)
\\
=&-\tint_{-\infty}^{\infty}\rd \bm{v}
f(\bm{x}, \bm{v},t)v_{k}(v_{l}-u_{l}(\bm{x},t)).
\label{eq:sigma}
\end{align}
Hence, by substituting 
$\xi(S_{k},t)$ and $\xi^{i}$ in Eq.~\eqref{eq:def_average_xi_on_Sk}
with
$-\frac{v_{k}(v_{l}-u_{l})}{|v_{k}|}$ and 
$-\frac{v_{k}^{i}(v_{l}^{i}-u_{l})}{|v_{k}^{i}|}$, respectively, 
it follows
\begin{align}
\nonumber
\tau^{\text{kin}}_{kl}(S_{k},t)
&=
\lim_{\delta t \to 0}\left[
-\frac{1}{S_{k}\delta t}\left<
\sum_{i\in\mathrm{fluid},\delta t}^{\text{crossing } S_{k}}
\frac{mv_{k}^{i}
\left(v_{l}^{i}-u_{l}(S_{k},t)\right)
}{
\left| v_{k}^{i}\right|}
\right>\right]
\\
\nonumber
&=
\lim_{\delta t \to 0}\left(
-\frac{1}{S_{k}\delta t}\left<
\sum_{i\in\mathrm{fluid},\delta t}^{\text{crossing } S_{k}}
\frac{mv_{k}^{i}v_{l}^{i}}{\left| v_{k}^{i}\right|}
\right>
+
u_{l}(S_{k},t)
\frac{1}{S_{k}\delta t}\left<
\sum_{i\in\mathrm{fluid},\delta t}^{\text{crossing } S_{k}}
\frac{mv_{k}^{i}}{\left| v_{k}^{i}\right|}
\right>\right)
\\
&=
-\lim_{\delta t \to 0}\frac{1}{S_{k}\delta t}\left<
\sum_{i\in\mathrm{fluid},\delta t}^{\text{crossing } S_{k}}
\frac{mv_{k}^{i}v_{l}^{i}}{\left| v_{k}^{i}\right|}
\right>
+
\rho u_{l}u_{k}(S_{k},t),
\label{eq:sigmasum}
\end{align}
where Eq.~\eqref{eq:massfluxsum} 
is used in the final equality. Note that the second term 
in the rightmost-HS can be obtained by
\begin{equation}
    \rho u_{l} u_{k}(S_{k},t)
    =
    \frac{\rho u_{l}(S_{k},t) \cdot \rho u_{k}(S_{k},t)}
    {\rho(S_{k},t)}
    \label{eq:advectionsum}
\end{equation}
using Eqs.~\eqref{eq:massfluxsum} and \eqref{eq:averagedvelocity},
which correspond to the advection term in the macroscopic 
momentum conservation 
in the Navier-Stokes equation%
~\eqref{eq:NS_macro}. 
%
By subtracting $\rho u_{l} u_{k}(S_{k},t)$ 
from the rightmost-HS and leftmost-HS of Eq.~\eqref{eq:sigmasum}, it follows
\begin{equation}
\tau^{\text{kin}}_{kl}(S_{k},t) - \rho u_{l} u_{k}(S_{k},t)
= 
-\lim_{\delta t \to 0}\frac{1}{S_{k}\delta t}\left<
\sum_{i\in\mathrm{fluid},\delta t}^{\text{crossing } S_{k}}
\frac{mv_{k}^{i}v_{l}^{i}}{\left| v_{k}^{i}\right|}
\right>,
\label{eq:sigmasum-rhovv}
\end{equation}
meaning that the microscopic total momentum transfer in the RHS corresponds to the stress minus the advection term in the LHS. 
Technically, the summation in the RHS of Eq.~\eqref{eq:sigmasum-rhovv}
is calculated during the MD simulation, and as the post process, the stress 
$\tau^{\text{kin}}_{kl}(S_{k},t)$
is obtained by adding the advection term $\rho u_{l}u_{k}$ to the total microscopic momentum transfer as
\begin{align}
\tau^{\text{kin}}_{kl}(S_{k},t) 
&= \left[\tau^{\text{kin}}_{kl}(S_{k},t) - \rho u_{l} u_{k}(S_{k},t)\right]
+ \rho u_{l}u_{k}(S_{k},t), 
\label{eq:sigmasum-prac}
\end{align}
where the advection term is  calculated by the dividing 
$\rho u_{l}(S_{k},t) \cdot \rho u_{k}(S_{k},t)$
by the density $\rho (S_{k},t)$ as in Eq.~\eqref{eq:advectionsum}: all  obtained also as the post process.

%
%
The relation between the macroscopic 
variables in Eqs.~\eqref{eq:mass_conserv_macro} and
\eqref{eq:NS_macro} corresponding microscopic 
expressions are summarized in TABLE~\ref{tab:table1}.
\begin{table*}[t]
\caption{\label{tab:table1} 
Microscopic expressions for the calculation of 
the corresponding macroscopic properties defined 
as the average on bin face $S_{k}$ in steady-state
systems. The top four properties can be directly 
calculated from steady-state systems through the 
MoP procedure, whereas the others below are derived 
from the four.
}
%
\begin{ruledtabular}
\begin{tabular}{ccc}
macroscopic property  &
microscopic expression & 
corresponding equation(s)
\\ \hline
$\rho(S_{k},t)$
&
$\ds  \lim_{\delta t \to 0} \frac{1}{S_{k}\delta t}\left<
\sum_{i\in\mathrm{fluid},\delta t}^{\text{crossing } S_{k}}
\frac{m}{\left| v_{k}^{i}\right|}
\right>$
 & 
Eq.~\eqref{eq:densitysum} 
\\
$\rho u_{l}(S_{k},t)$
& 
$\ds \lim_{\delta t \to 0}
\frac{1}{S_{k}\delta t}\left<
\sum_{i\in\mathrm{fluid},\delta t}^{\text{crossing } S_{k}}
\frac{mv_{l}^{i}}{\left| v_{k}^{i}\right|} 
\right>$
 &  
Eq.~\eqref{eq:massfluxsum}
\\
$\tau_{kl}^\mathrm{int}(S_{k},t)$
&
    $\ds -\frac{1}{S_{k}}
    \angb{
    \sum_{(i,j)\in \mathrm{fluid}}^{\text{across\ } S_{k}} 
    F^{ij}_{l} \frac{r^{ij}_{k}}{|r^{ij}_{k}|}
    }$
    &
    Eq.~\eqref{eq:stress_int}
\\
$\tau_{kl}^\mathrm{kin}(S_{k},t) - \rho u_{l}u_{k}(S_{k},t)$
&
$\ds 
-  \lim_{\delta t \to 0} \frac{1}{S_{k}\delta t}\left<
\sum_{i\in\mathrm{fluid},\delta t}^{\text{crossing } S_{k}}
\frac{mv_{k}^{i}v_{l}^{i}}{\left| v_{k}^{i}\right|}
\right>$
&
Eq.~\eqref{eq:sigmasum-rhovv}
\\
\hline
$\ds u_{l} = \frac{\rho u_{l}}{\rho}$
&
-
&
Eq.~\eqref{eq:averagedvelocity}
\\
$\rho u_{l}u_{k}$
&
-
&
Eq.~\eqref{eq:advectionsum}
\\
$\tau_{kl}^\mathrm{kin} =
\left( \tau_{kl}^\mathrm{kin} - \rho u_{l}u_{k} \right) 
+ \rho u_{l}u_{k}$
&
-
&
Eq.~\eqref{eq:sigmasum-prac}
\\
$\tau_{kl}=\tau_{kl}^\mathrm{kin}+\tau_{kl}^\mathrm{int}$
&
-
&
Eq.~\eqref{eq:tau=taukin+tauint}
\end{tabular}
\end{ruledtabular}
\end{table*}

In practice, within the framework of MD, 
$\delta t\ (\to 0)$ must be replaced
by a small non-zero time step of $\Delta t$ for the numerical integration.
Upon this procedure without this limit, 
we have to assume the following: 
1) the change of the distribution function $f(\bm{x}, \bm{v},t)$ 
within the distance range of $| \bm{v}|\delta t$ is negligibly
small,  and 
2) the values of $v^{i}_{k}$ and $v^{i}_{l}$ 
upon `crossing' should be properly evaluated based on the position update procedure of particles depending on the time integration scheme.
For the velocity Verlet method, which is applied in the numerical test in Sec.~\ref{sec:Test}, we adopted $\bm{v}^{i} \equiv \frac{\bm{x}^{i}(t+\Delta t)-\bm{x}^{i}(t)}{\Delta t}$ using the positions $\bm{x}^{i}(t)$ and $\bm{x}^{i}(t+\Delta t)$ of fluid particle $i$ at time $t$ and $t+\Delta t$ before and after crossing the bin face 
to avoid the discrepancy of the mass flux by the MoP calculation and by the position update.


Note that Eq.~\eqref{eq:sigmasum-rhovv} without the limit 
$\delta t \to 0$
is the same as the RHS of Eq.~\eqref{eq:MoP_kin}, which 
simply sums up the momentum transfer across the bin face $S_{k}$
with a sign function $\frac{v_{k}^{i}}{|v_{k}^{i}|}$.
%
Hence, if one locates a control volume with a closed surface consisting of the MoP bin faces, 
then the momentum conservation is strictly satisfied
with Eq.~\eqref{eq:sigmasum-rhovv}.
Different choices are indeed possible to determine
the advection 
term
$\rho u_{l} u_{k}(S_{k},t)$ in Eq.~\eqref{eq:advectionsum} 
to separate the stress $\tau_{kl}^\mathrm{kin}(S_{k},t)$ from 
$\tau_{kl}^\mathrm{kin}(S_{k},t) - \rho u_{l}u_{k}(S_{k},t)$
by Eq.~\eqref{eq:sigmasum-prac}, and this may sound that the definition of $\tau_{kl}^\mathrm{kin}(S_{k},t)$ is not unique. 
However;  
by setting $l=k$ in  Eq.~\eqref{eq:massfluxsum}, 
the surface normal mass flux is evaluated 
as the simple sum of the mass passage with a sign function 
$\frac{v^{i}_{k}}{|v^{i}_{k}|}$, and this strictly 
satisfies the mass conservation 
in Eq.~\eqref{eq:mass_conserv_macro_vol_surf}, meaning that one can choose a unique definition 
of $\rho u_{l} u_{k}(S_{k},t)$ that simultaneously satisfies the macroscopic mass and momentum conservation. 
%
\par
Another point to be noted is that the final forms in Eqs.~\eqref{eq:densitysum}, \eqref{eq:massfluxsum} 
and \eqref{eq:sigmasum-rhovv} are formally equivalent to the MoP 
expressions by \citet{Daivis1996}, which were derived for a quasi-1-dimensional flow through the expressions of the time derivative of the fluxes in a control volume with the Fourier transform, and were in principle applicable for the average on an infinite plane  under a periodic boundary condition.
On the other hand, our non-flux-based derivation 
with a 
definition of physical properties averaged on a face through the VDF
enables the calculation of physical properties on 
a finite area. In addition, taking advantage of this non-flux-based feature, one can calculate, for instance, the velocity component $u_{l}$ on a bin face $S_{k}\ (l\neq k)$ tangential to the velocity component by Eq.~\eqref{eq:averagedvelocity}.
This point will be discussed more in detail with a quasi-1D Couette-type flow in Sec.~\ref{subsec_Q1D} as an example.
%
%

\label{sec:Method}
\section{Numerical Test}
\label{sec:Test}
The extended MoP was tested through the calculation 
of the density, macroscopic mean velocity and 
stress distributions in two systems with 
a Lennard-Jones (LJ) fluid
: a quasi-1D Couette-type flow and a quasi-2D shear flow with solid-liquid-vapor contact lines. 
Note that both systems are in steady state and we applied time average instead of ensemble average.
\subsection{Quasi-1D Couette-type flow}
\label{subsec_Q1D}
\begin{figure}[t]
  \begin{center}
    \includegraphics[width=\linewidth]{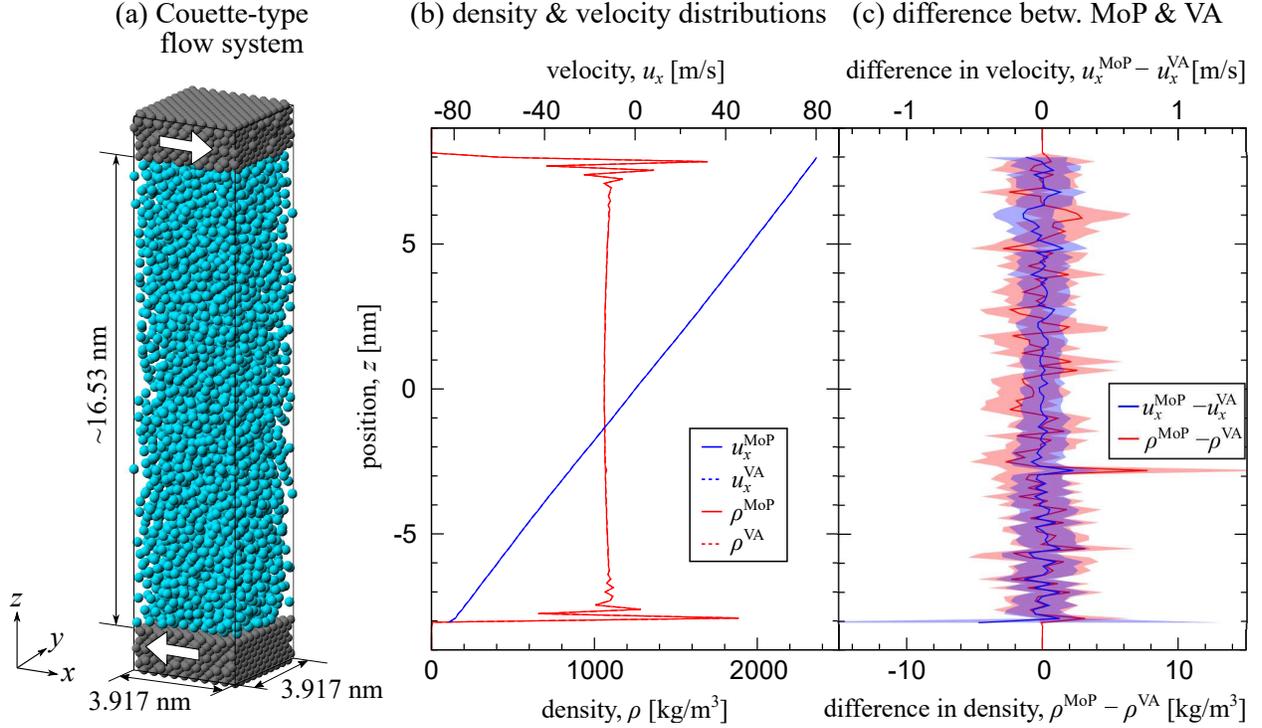}
  \end{center} 
  \caption{\label{fig:fig1}
(a) Quasi-1D Couette-type flow system of a Lennard-Jones liquid 
confined between two solid walls.
(b) Distributions of density $\rho$ and velocity $u_{x}$ calculated by the proposed Method of Plane (MoP) and the volume average (VA). 
Solid and dashed lines denote the results of MoP and VA, respectively while the two lines almost overlap 
in this scale. 
(c) Difference between MoP and VA regarding 
density $\rho^{\text{MoP}}-\rho^{\text{VA}}$ 
and velocity $u_{x}^{\text{MoP}}-u_{x}^{\text{VA}}$ with their error bars depicted with semi-transparent areas around the average.
}
\end{figure}

Figure~\ref{fig:fig1}~(a) shows the MD simulation system of a 
quasi-1D Couette-type flow, where the basic setup is a standard one similar to our 
previous study.\cite{Ogawa2019, Oga2019}
The two parallel solid walls were fcc crystals and 
every pair of the nearest neighbors in the walls was bound through a harmonic potential
$\Phi_{\text{h}}(r)=\frac{k}{2}(r-r_\text{eq})^2$, with $r$ being the interparticle distance, $r_\text{eq}=0.277$~nm, and $k=46.8$~N/m.
Interactions between fluid particles and between fluid and solid particles were modeled by a 12-6 LJ potential
$\Phi^\mathrm{LJ}(r_{ij}) =  4\epsilon_{ij} \left[ \left(\frac{\sigma_{ij}}{r_{ij}}\right)^{12}-
\left(\frac{\sigma_{ij}}{r_{ij}}\right)^{6} \right] $,
where $r_{ij}$ was the distance between the particles $i$ and $j$, while $\epsilon$ and $\sigma$ denoted the LJ energy and length parameters, respectively.
This LJ interaction was truncated at a cut-off distance of $r_\mathrm{c}=3.5 \sigma$ and quadratic functions were added so that the potential and interaction force smoothly vanished at $r_\mathrm{c}$.~\cite{Nishida2014}
We used the following parameters for fluid-fluid (ff) and solid-fluid (sf) interactions:
$\sigma_{\text{ff}}=0.340$~nm, $\epsilon_{\text{ff}}=1.67\times10^{-21}$~J, 
$\sigma_{\text{sf}}=0.345$~nm, $\epsilon_{\text{sf}}=0.646\times10^{-21}$~J.
The atomic masses of fluid and solid particles were $m_{\text{f}}=39.95$~u and $m_{\text{s}}=195.1$~u, respectively.
Finally, the equations of motion were integrated using the velocity-Verlet algorithm, with a time step $\Delta t$ of 5 fs.

The periodic boundary condition was set in the $x$ and $y$-directions, 
and 4000 LJ particles were confined between two parallel solid walls consisting of the fcc crystal located on the bottom and top sides of the calculation cell, 
which directed (001) and (00$-1$) planes normal to the $z$-direction.
Both had eight layers so that the possible minimum distance between the fluid particle 
and the solid particle in the outmost layer was longer than the cutoff distance.
The relative positions of the solid particles in the outmost layers of each base crystal were fixed 
and the temperature of those in the second outermost layers was controlled at a control temperature of 100~K by using the standard Langevin thermostat.~\cite{Blomer1999}
The system was first equilibrated for 10 ns using the top wall as a piston with a control pressure of 4~MPa without shear so that a quasi-1D system with a LJ liquid confined between fcc solid walls was achieved.
After the equilibration, further relaxation run to achieve a steady shear flow was carried out for 10~ns by moving the particles in the outmost layers of both walls with opposite velocities of $\pm$100~m/s in the $x$-direction, using the top wall as a piston with a control pressure of 4~MPa. 
Finally, steady shear flow simulation was carried out, keeping their $z$-position constant at the average position during the 2nd relaxation run, where the system pressure resulted in 3.61~MPa.

We tested the MoP expression (TABLE~\ref{tab:table1}) in the steady 
state,
where the local density, 
velocity, 
advection term, 
and stress 
were obtained as the time-average of 200~ns 
on a grid with $x$-normal bin faces with a height $\Delta z=0.150$~nm and $z$-normal ones with a width $\Delta x=0.145$~nm. 
Assuming that the system is quasi-1D, 
the distribution in the $x$-direction was 
averaged for bins with identical 
$z$-positions.  
%
For a comparison, we also obtained the 
density and velocity distributions based on
a standard volume average (VA), where
the time-average in equally divided bin 
volumes parallel to the solid wall with a 
height of 0.15~nm were calculated.
%

Figure~\ref{fig:fig1}~(b) shows the distributions of 
density $\rho$ and macroscopic velocity in the $x$-direction $u_{x}$ 
calculated by the proposed MoP and standard VA as a reference.
Note that these distributions by the MoP 
can be calculated both on $x$-normal bin faces 
and on $z$-normal ones as 
shown later, 
while only the distributions obtained on $x$-normal 
bin faces are shown as the MoP results here. 
Overall, the MoP well 
reproduced the results by the VA, and 
the two lines almost overlap in this scale. 
Regarding the density distribution, 
except near the walls where layered structures  
are observed, 
bulk liquid with almost constant density was formed. 
Note that the bulk density was not completely constant because the temperature was not constant due to the viscous heat dissipation 
induced by the extreme shear imposed on this system.
The shear velocity profiles 
are linear throughout almost all the liquid 
part except in layered structures, which 
can be understood by the change of local 
viscosity there. 
%
%
The density and velocity differences between MoP and 
VA are shown in Fig.~\ref{fig:fig1}~(c). 
The density difference was within 10~$\text{kg}/\text{m}^3$, which is less than 
1~\%  
of the bulk density, and the velocity difference was also within 0.5~m/s, showing that proposed 
MoP can extract the density and velocity distribution 
consistent with VA.
%
%
%
\begin{figure}[t]
  \begin{center}
    \includegraphics[width=0.9\linewidth]{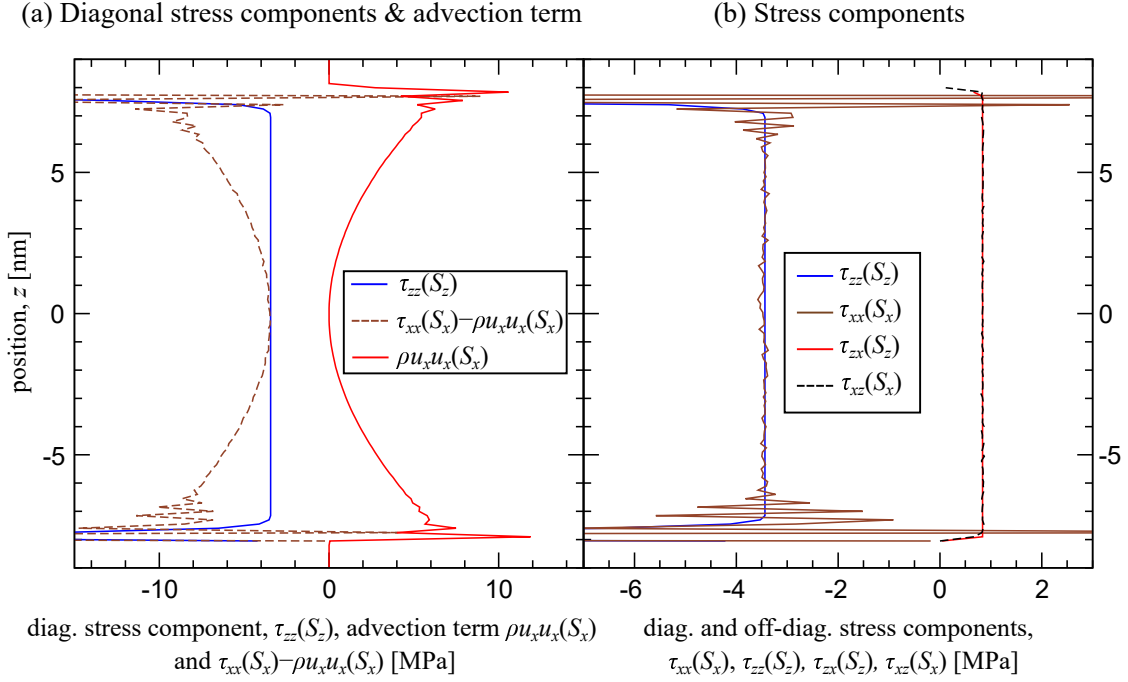}
  \end{center} 
  \caption{\label{fig:fig2}
Distributions of (a) the diagonal stress component  
$\tau_{zz} (S_{z})(\equiv \tau_{zz}^\mathrm{int}+\tau_{zz}^\mathrm{kin})$, 
advection term $\rho u_{x}u_{x}(S_{x})$ and 
$\tau_{xx}(S_{x})-\rho u_{x}u_{x}(S_{x}) [\equiv \tau_{xx}^\mathrm{int}+(\tau_{xx}^\mathrm{kin}-\rho u_{x}u_{x})]$, and (b) diagonal and off-diagonal stress components 
$\tau_{xx}(S_{x})$, $\tau_{zz}(S_{z})$, $\tau_{zx}(S_{z})$ and $\tau_{xz}(S_{x})$. 
}
\end{figure}
\par
Figure~\ref{fig:fig2}~(a) shows the distributions of 
$\tau_{zz} (\equiv \tau_{zz}^\mathrm{int}+\tau_{zz}^\mathrm{kin})$, 
$\tau_{xx}-\rho u_{x}u_{x} [\equiv \tau_{xx}^\mathrm{int}+(\tau_{xx}^\mathrm{kin}-\rho u_{x}u_{x})] $
and $\rho u_{x}u_{x}$, where the first two were directly 
obtained with simple addition based on Eqs.~\eqref{eq:stress_int} and 
\eqref{eq:sigmasum} as also listed in the top part of 
TABLE~\ref{tab:table1}, while $\rho u_{x} u_{x}$ was obtained 
from the density $\rho$ and velocity $u_{x}$. Note that 
$\tau_{zz} - \rho u_{z}u_{z}$ is shown as $\tau_{zz}$ because $u_{z}$ is equal to zero in the whole area of the present system.
Also note that the calculation of $\tau_{kl}^\mathrm{int}$ 
in Eq.~\eqref{eq:stress_int} was the same as in 
equilibrium systems without macroscopic flow. 
As clearly observed, $\tau_{xx}-\rho u_{x}u_{x}$ including 
the advection and $\tau_{zz}$ are different away from the 
solid walls, indicating that the flow effect should be 
removed to properly evaluate the fluid stress. 
Figure~\ref{fig:fig2}~(b) 
displays the distributions of the stress component $\tau_{xx}$, 
$\tau_{zz}$, $\tau_{zx}$ and $\tau_{xz}$, 
where $\tau_{zz}$  and $\tau_{zx}$ were 
calculated on $z$-normal bins whereas the others 
were obtained on $x$-normal bins. As explained above, 
the stress 
value $\tau_{xx}$ was calculated by adding 
$\rho u_{x} u_{x}$ to $\tau_{xx}-\rho u_{x}u_{x}$ whereas 
the advection terms for the others can be neglected 
considering $u_{z}=0$.
In the bulk region 
sufficiently away from the walls, $\tau_{xx}=\tau_{zz}$ and $\tau_{zx}=\tau_{xz}$ 
are satisfied as expected from the solution of a laminar Couette flow, and 
the former indicates that the stress $\tau_{xx}$ is adequately calculated by 
the proposed 
MoP with the resulting value $-\tau_{xx}(=-\tau_{zz})$ equal to the external
pressure value of 3.61~MPa. The wall-tangential diagonal stress $\tau_{xx}$ 
fluctuates near the walls as typically observed also in equilibrium systems,~\cite{Nishida2014,Yamaguchi2019} 
because of the layered structure of the liquid 
as displayed in the density distribution in Fig.~\ref{fig:fig1}~(b). On the other hand, $\tau_{zz}$ was constant except near the walls, where the solid-liquid (SL) interaction acts as the external force on the liquid. 
Regarding the off-diagonal components $\tau_{zx}(=\tau_{xz})$, they were 
constant except just around the walls, where friction from 
the solid is included in the force balance even in the laminar flow.
%
%
%
%
\par
\begin{figure}[t]
  \begin{center}
    \includegraphics[width=0.9\linewidth]{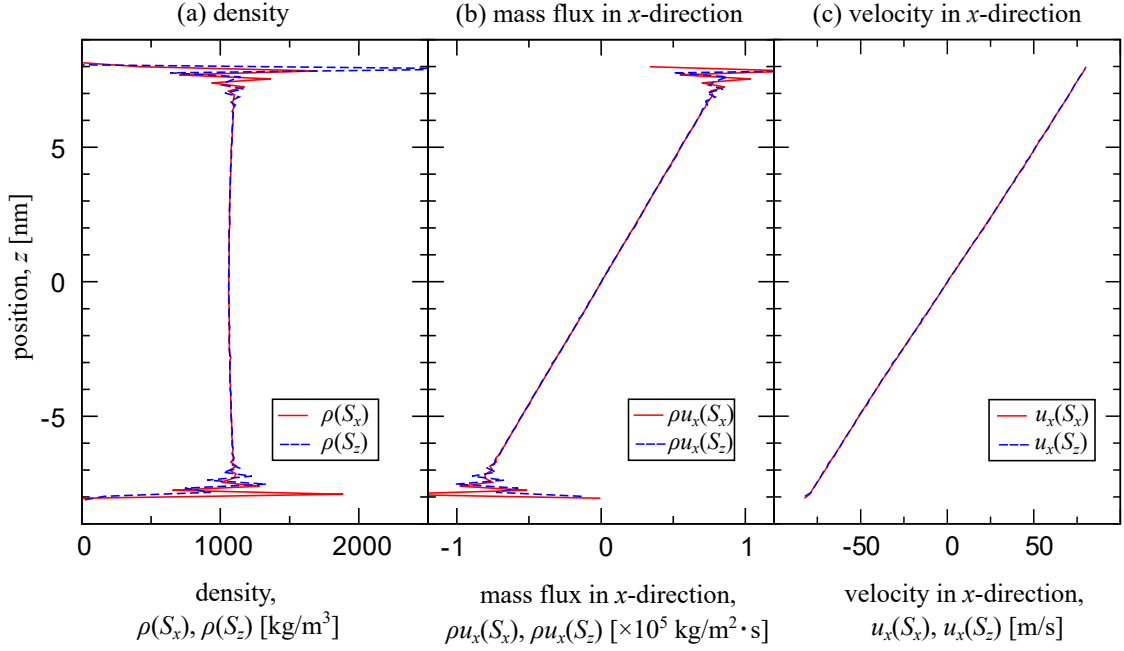}
  \end{center} 
  \caption{\label{fig:fig3}
  Comparison of the time-averaged distributions of the (a) density $\rho$ (b) mass flux $\rho u_{x}$ and (c) velocity $u_{x}$ averaged on $x$-normal and $z$-normal bin faces $S_{x}$ and $S_{z}$, respectively.
}
\end{figure}

In addition to the normal velocity component $u_{k}$ on the MoP plane $S_{k}$, the calculation of $u_{l} \ (l\neq k)$ tangentially to $S_{k}$ is needed for the separation of $\tau_{kl}^\mathrm{kin}(S_{k})-\rho u_{l}u_{k}(S_{k})$ in Eq.~\eqref{eq:sigmasum-prac} to properly define the stress in general flows with $u_{l}\neq 0$ and $u_{k}\neq 0$.
Including this tangential velocity, we compared the distributions of the density $\rho$, the mass flux $\rho u_{x}$ and the velocity $u_{x}$  averaged on $x$-normal and $z$-normal bin faces as another numerical test in the present system in Fig.~\ref{fig:fig1}. More concretely, the density  $\rho (S_{x})$ and $\rho (S_{z})$ averaged on $x$-normal and $z$-normal bin faces $S_{x}$ and $S_{z}$, respectively were calculated by Eq.~\eqref{eq:densitysum} with setting $k=x$ and $k=z$, whereas the macroscopic mass flux $\rho u_{x}(S_{x})$ and $\rho u_{x}(S_{z})$ were obtained by Eq.~\eqref{eq:massfluxsum} with $l=x$ on $S_{x}$ and $S_{z}$, 
respectively. With these definitions, $u_{x}(S_{x}) \equiv \frac{\rho u_{x}(S_{x})}{\rho(S_{x})}$ on $S_{x}$ as well as $u_{x}(S_{z}) \equiv \frac{\rho u_{x}(S_{z})}{\rho(S_{z})}$ on $S_{z}$ can be obtained as in Eq.~\eqref{eq:averagedvelocity}.
Note that in the present laminar flow system with $u_{z}=0$, the calculation of $u_{x}$ is practically not needed for the stress separation 
for $\tau_{zz}^\mathrm{kin}$, $\tau_{zx}^\mathrm{kin}$ and $\tau_{xz}^\mathrm{kin}$ in Eq.~\eqref{eq:sigmasum-prac}.
%
%
%
%
%
%
Figure~\ref{fig:fig3} shows the distributions of the (a) density $\rho$, (b) mass flux $\rho u_{x}$, and (c) velocity $u_{x}$ defined on $x$-normal 
and $z$-normal bin faces, 
in the system in Fig.~\ref{fig:fig1}, 
where the values averaged on each bin face of $x$-normal and $z$-normal  are plotted with setting the $z$-position at the center of each bin face, respectively,  \ie they are staggered by $\Delta z/2$.
In the bulk, $\rho$, $\rho u_{x}$ and resulting $u_{x}$ averaged 
on bin faces with different normal directions 
agreed well, indicating that the separation of the stress and advection terms in Eq.~\eqref{eq:sigmasum-prac} is possible 
with the velocity values properly evaluated by the proposed method.
The difference seen around the top and bottom is due to the 
layered structures around the two walls, \ie the values on 
$S_{z}$ are the average on a surface parallel to the layered structure
whereas those on $S_{x}$ are the average across the layers.
%
\subsection{Quasi-2D shear flow with 
solid-liquid-vapor contact lines}
\begin{figure}[t]
  \begin{center}
    \includegraphics[width=0.8\linewidth]{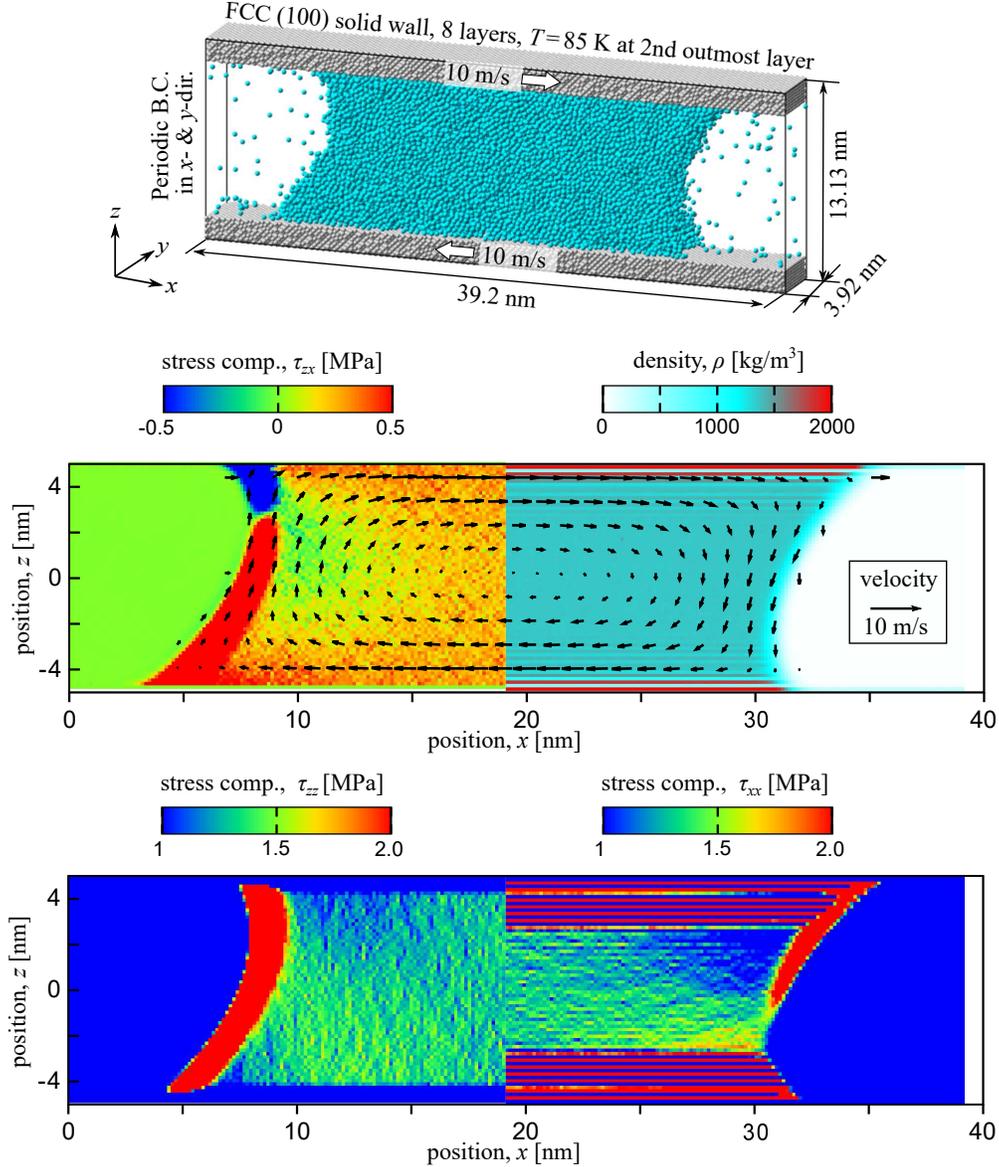}
  \end{center} 
  \caption{\label{fig:fig4}
Top: Quasi-2D Couette-type flow system of a Lennard-Jones liquid 
confined between two solid walls.
Middle: Distributions of density $\rho$, velocity $\bm{u}$, and off-diagonal stress component $\tau_{zx}$. Black arrow denotes the macroscopic velocity calculated by the proposed Method of Plane.
Bottom: Distributions of diagonal stress components $\tau_{xx}$ and $\tau_{zz}$.
}
\end{figure}
%
%
%
The top panel of Fig.~\ref{fig:fig4} shows the MD simulation system of a quasi-2D Couette-type flow, where 
the basic setups 
are the same as in the quasi-1D system. 
%
The periodic boundary condition was set in the $x$- and $y$-directions, and 20,000 LJ particles were 
confined between two parallel solid walls with a distance about 10.4~nm and a dimension of $x \times y = 3.92\times 39.2$~nm$^{2}$ 
so that the LJ fluid may form two quasi-2D menisci with contact lines on the walls upon the 
preliminary equilibration at a control temperature $T=85$~K without shear. The static contact angle on both top and bottom walls were about 57 degrees.
After the equilibration, further relaxation run to 
achieve a steady shear flow with asymmetric menisci
were carried out for 
10~ns by moving the particles in the outmost layers 
of both walls with opposite velocities of $\pm$10~m/s 
in the $x$-direction. 
\par
After the relaxation run, the density, velocity and stress distributions were obtained by the present MoP expression in 
the steady state with the time-average of 500~ns 
on $x$-normal bin faces with a length of $\Delta z=0.149$~nm 
and $z$-normal ones with a length of $\Delta x=0.150$~nm.
\par
The middle panel of Fig.~\ref{fig:fig4} shows the 
distributions of density $\rho$ calculated on the $x$-normal bin faces, velocity vector with components calculated on each bin face corresponding the component direction, and a stress component $\tau_{zx}$, where those for 
$\rho$ and $\tau_{zx}$ are displayed only for the 
half of the system with respect to the center of 
mass of the fluid considering the symmetry.
A clockwise caterpillar like flow is clearly captured by the present method, where the shear stress $\tau_{zx}$ distribution in the liquid phase shows the non-uniformity of the viscous stress. 
The strong tensile stress seen in the $\tau_{zx}$ distribution around the LV interfaces is due to the LV interfacial tension.
The bottom panel of Fig.~\ref{fig:fig4} shows the distributions of diagonal stress components $\tau_{xx}$ and  $\tau_{zz}$. Layered structures are observed for $\tau_{xx}$ 
near the SL interfaces due to the adsorption layers in the density distribution. The relation between the density layers and stress distribution 
near the solid walls are qualitatively the same as in our previous study of static droplet:~\cite{Yamaguchi2019} negative stress was seen in the adsorption layers, \ie the adsorption layers were compressed 
whereas tensile stress appeared between the layers. 
%
%

Indeed, these apparent flow features can be qualitatively visualized by another methods such as atomic stress~\cite{Thompson2009}, but the present method provides the distributions of physical properties defined on a surface 
establishing a direct link with the conservation laws for arbitrary local volume as described in \textit{Introduction}, and is generally applicable to a wide range of nanoscale systems with liquid flow. One of our future research targets is dynamic wetting~\cite{Hizumi2015,Omori2019,Thalakkator2020}, for which we plan to examine the mechanical balance exerted on the fluid around a CV set around the moving contact line in Fig.~\ref{fig:fig4}. 
Through the comparison with the static  case,~\cite{Yamaguchi2019} this would enable the analysis of advancing and receding contact angle from a mechanical point of view. 

\section{Concluding remarks}
In this work, we showed a calculation method of local stress tensor applicable to non-equilibrium MD systems based on the Method of Plane (MoP). 
 From the relation between the macroscopic velocity distribution function and the microscopic molecular passage across a fixed control plane, we derived a basic equation to connect the macroscopic field variable and the microscopic molecular variable. 
Based on the connection, we derived a method to calculate the basic properties of the macroscopic momentum conservation law including the density, velocity and momentum flux as well as the interaction and kinetic terms of the stress tensor defined on a surface with a finite area. 
Any component of the streaming velocity can be obtained on a control surface, 
which enables the separation of the kinetic momentum flux into the advection and stress terms in the framework of the MoP.
We verified the present method through the extraction of the density and velocity distributions by volume average (VA) and the MoP in a quasi-1D steady-state Couette flow system, seeing that the stress tensor distribution by the MoP satisfies the solution of a laminar Couette flow in the bulk, indicating that the flow effect, \ie the advection term, was removed to evaluate stress properly.
Furthermore, we showed the density, velocity, and stress tensor distributions by the MoP even in a quasi-2D steady-state system with a moving contact line. 
In our method as opposed to VA, 
the density, mass and momentum fluxes 
are defined on a surface, which is essential 
to have consistency with the 
conservation laws 
in dynamic systems.
\begin{acknowledgments}
H.K., T.O. and Y.Y. are supported by JSPS KAKENHI Grant 
Nos. JP20J20251, JP18K03929, and JP18K03978, Japan, respectively. 
Y.Y. is also supported by JST CREST Grant No. JPMJCR18I1, Japan.
\end{acknowledgments}
%
%
\vspace{5mm}
\par \noindent
\textbf{DATA AVAILABILITY}
\newline
\par
The data that support the findings of this study are available from the corresponding author
upon reasonable request.
\bibliography{reference}
%
%
\end{document}